\documentstyle[preprint,aps,12pt]{revtex}
\input{epsf}

\newcommand{\be}{\begin{equation}}
\newcommand{\ee}{\end{equation}}
\newcommand{\omn}{\bar{\Omega}_n}

\newcommand{\sumk}{\sum_{k=0}^{\infty}}

\newcommand{\mur}{\bar{\mu}}

\tighten

\begin{document}
\preprint{LA PLATA-TH 98/18}
\bibliographystyle{prsty}
\draft

\title{Finite density and temperature in hybrid bag models}

\author{M. De Francia \and H. Falomir}

\address{Departamento de F\'{\i}sica, Facultad de Ciencias Exactas,
Universidad Nacional de La Plata, \\ C.C. 67, 1900 La Plata,
Argentina}

\author{M. Loewe}

\address{Facultad de F\'{\i}sica, Pontificia Universidad Cat\'{o}lica de
Chile, \\ Casilla 306, Santiago 22, Chile}

\date{\today}

\maketitle


\begin{abstract}

We introduce the chemical potential in a system of two-flavored
massless fermions in a chiral bag by imposing boundary conditions
in the Euclidean time direction. We express the fermionic mean
number in terms of a functional trace involving the Green function
of the boundary value problem, which is studied analytically.
Numerical evaluations for the fermionic number are presented.

\end{abstract}


\section{Introduction}
\label{sec-introduccion}

Functional determinants of elliptic differential operators are useful for
describing one-loop effects in Quantum Field Theory and Statistical
Mechanics.

When the fields are constrained to live inside a bounded manifold in the
Euclidean space-time, boundary conditions become relevant in the
calculation of physical quantities.

This is the case of Hybrid Chiral bags, which are effective models
for confinement \cite{rhoreport,vento,zahed} (For a general review
on bag models see \cite{hadrons}). They share the characteristics
of two succesful approaches to describe baryons: bag models
\cite{mit,quiral} and Skyrme model \cite{skyrme,witten,adkins}.

According to the Cheshire Cat Principle (CCP)
\cite{rhoreport,nadkarni}, in these two-phase models, fermionic
degrees of freedom could be replaced by bosonic ones in a region
of space, physical quantities being insensitive to the position of
the limit of separation of the two phases. In $1+1$-dimensions the
CCP follows from the bosonization of fermionic fields
\cite{rhoreport}. In the $3+1$-case, topological quantities such
as the baryonic number at $T=0$ have a similar behavior
\cite{goldstone}, while for non-topological ones, as the total
energy at $T=0$, the CCP is approximately valid \cite{cheshire}.

In this paper we will study the mean fermionic number for a two
phase model at finite temperature and chemical potential. This
represents an extension of previous calculations done in the frame
of the MIT bag model \cite{mu-MIT}.

In Section~\ref{sec-setting} we will set the problem, expressing
the mean fermionic bag number as an adequate trace of the Green
function of the system. In doing this, the chemical potential is
introduced through the boundary conditions in the Euclidean time
direction \cite{chempot,mu-MIT}.

Section~\ref{sec-green} is devoted to the calculation of the Green
function for the Dirac operator obeying the above temporal
boundary condition, and spatial (chiral bag) boundary conditions
which relate the fermionic degrees of freedom in the inner phase
with the Skyrme field in a hedgehog configuration outside of the
bag.

In Section~\ref{sec-constructing} the expression of the mean
fermionic number is built. By means of the asymptotic Debye
expansion of Bessel functions we are able to split the mean
fermionic number into two pieces: a regular one, to be evaluated
numerically, and a possibly non-regular one, to be treated
analytically following the techniques presented in
Ref.~\cite{mu-MIT}.

In Section~\ref{sec-results} we show our numerical results, for
the total mean fermionic number including both the contributions
of the bag sector and of the skyrmion sector as in
Ref.~\cite{witten}.

Finally, in Section~\ref{sec-conclusion} the conclusions are presented.


\section{The hybrid bag model}
\label{sec-setting}

We will be interested in the study of the mean fermionic number
for a two phase model: a two-flavored massless fermionic field
confined inside a static sphere of radius $R$ and a hedgehog
Skyrme field filling the exterior sector. These two phases are
linked by the spatial chiral boundary conditions, which will be
introduced in the following.

The fermionic field inside the bag $\Sigma \otimes [0,1]$, $0 \leq
|\vec{x}|,t \leq 1$ is described by the Dirac operator
\be
D(\beta,R) = \left[ \frac{i}{\beta} \gamma^0 \partial_t +
\frac{i}{R} \vec{\gamma} \cdot \vec{\nabla} \right] \otimes {\cal
I}_I \label{eq-diffop}\ee
Here ${\cal I}_I$ denotes the identity in the flavor (isospin) space, and
$\beta$ stands for the inverse temperature.

The external phase is described by a Skyrme model
\cite{skyrme,skyrme2}, whose Lagrangian is given by
\be
{\cal L} = \frac{1}{16} F_\pi^2 {\rm Tr} \left( \partial_\mu U
\partial^\mu U^\dagger \right) + \frac{1}{32 e^2} {\rm Tr}
\left[\left(  \partial_\mu U \right) U^\dagger , \left(
\partial_\nu U \right) U^\dagger \right]^2, \label{eq-lagr-skyrme}\ee
where the scalar field $U(x)$ takes values in the $SU(2)$ group.

When the whole space is filled by $U(x)$, a topological stable
classical solution is given by the hedgehog configuration
\be
U_0 = e^{i \theta (r) \left( \vec{\tau} \cdot \vec{x} \right)},
\label{hedgehog} \ee
where the profile of the skyrmion $\theta
(r)$ is the chiral angle. In this pure skyrmionic model, the
imposition of the boundary conditions
\[
\theta (r = 0) = \pi, \qquad \theta (r \rightarrow \infty) \rightarrow 0
\]
leads to a topological charge (winding number) ${\cal B} = 1$, which is
identified with the baryonic number of the skyrmion \cite{witten,adkins}.
The profile of the chiral angle, $\theta(r)$, is numerically determined by
minimizing the classical energy of the soliton \cite{adkins,rho}.

When the skyrmion fills only the exterior of the bag, its contribution to
the fermionic number, $N_{\rm Sk}$, is given by the integral of the
topological charge density over that region, giving
\be
N_{\rm Sk} = \frac1{\pi} \left( \theta - \sin\theta \cos\theta
\right) \label{numero-skyrmion} \ee as discussed in Ref.
\cite{witten}.

In the hybrid bag model, fermionic and skyrmionic phases are
connected through the spatial boundary conditions
\be
B \psi(t,\vec{x}) = 0 \quad {\rm for}\, |\vec{x}|=1 \ee
where
\be
B = {\cal I}_4 \otimes {\cal I}_I + \left( i \not n \otimes {\cal
I}_I \right) e^{-i \theta \gamma^5 \otimes \left( \vec{\tau} \cdot
\vec{n}\right)} , \label{eq-boundop}\ee
being $\theta=\theta(R)$ the value of the skyrmion profile at the surface
of the bag.

Since we are dealing with fermions, these fields must be
antiperiodic in the Euclidean time direction,

\be
\psi (1,\vec{x}) = - \psi (0,\vec{x}). \ee

The Grand Canonical partition function is given by
\be
\Xi (T,R,\mu) = e^{-\beta G(T,R,\mu)} \sim {\rm Det} \left[
D(\beta,R) - i \mu \gamma^0 \right]_{\rm BC}, \ee
where ``BC" means
that the differential operator is defined on a space of functions
satisfying both spatial and temporal boundary conditions. In the
above expressions, $\mu$ stands for the chemical potential.

As discussed in Ref.~\cite{mu-MIT} the mean fermionic number can
be expressed as
\begin{equation}
\langle N_{\rm bag} \rangle (\beta, R,\mu;\theta)= {\displaystyle
-\frac{\partial G}{\partial \mu }(\beta ,R,\mu ;\theta
)}={\displaystyle \frac1\beta} Tr\{-i \left( \gamma ^0 \otimes
{\cal I}_I \right) k (t,x;t^{\prime },x^{\prime })\},
\label{numero}
\end{equation}
where $k(t,\vec{x};t^\prime , \vec{x^\prime})$ is the Green
function of the problem
\begin{eqnarray}
  D(\beta,R) k(t,\vec{x};t^\prime , \vec{x^\prime}) & = &
  \delta^{(3)} (\vec{x}-\vec{x^\prime}) \, \delta (t-t^\prime) {\cal I}_8,
  \qquad {\rm for} \, \vec{x},\vec{x^\prime} \in \Sigma,
  \label{eq-eqdiff} \\
  B k(t,\vec{x};t^\prime , \vec{x^\prime}) & = & 0,  \qquad \qquad
  {\rm for}\, \vec{x}
  \in \partial\Sigma, \label{eq-bcspatial} \\
  k(1,\vec{x};t^\prime , \vec{x^\prime}) & + &
  e^{-\mu \beta} k(0,\vec{x};t^\prime , \vec{x^\prime})= 0.
\label{eq-bctemp} \end{eqnarray}

The total mean fermionic number for the hybrid chiral bag model is
defined as
\be
\langle N \rangle (\beta, R, \mu ;\theta) = \langle N_{\rm bag}
\rangle (\beta, R, \mu ;\theta) + N_{Sk} (\theta) \ee

One can apply discrete transformations to these equations to
obtain symmetries of $\langle N_{\rm bag} \rangle$. In fact,
\be
k_1(t,\vec{x};t^\prime , \vec{x^\prime}) = - \gamma^5
k(t,\vec{x};t^\prime , \vec{x^\prime}) \gamma^5 \ee is the Green
function of the problem with $\theta \rightarrow \theta + \pi$ in
the spatial boundary condition, equation (\ref{eq-bcspatial}).
Then, from equation (\ref{numero}), it is easy to establish that
\be
\langle N_{\rm bag} \rangle (\beta, R, \mu ;\theta + \pi) =
\langle N_{\rm bag} \rangle (\beta, R, \mu ;\theta)
\label{eq-sim1}\ee

Taking into account (\ref{numero-skyrmion}) and (\ref{eq-sim1}),
the total mean fermionic number satisfies
\be
\langle N \rangle (\beta, R, \mu ;\theta + \pi) = 1 + \langle N
\rangle (\beta, R, \mu ;\theta) .\ee

Moreover,
\be
k_2(t,\vec{x};t^\prime , \vec{x^\prime}) =  \gamma^0 \gamma^5
k(1-t,\vec{x};1-t^\prime , \vec{x^\prime}) \gamma^5 \gamma^0 \ee
is the Green function of the problem with $\theta \rightarrow -
\theta$ in the spatial boundary condition and $\mu \rightarrow
-\mu$ in the Euclidean time boundary condition (\ref{eq-bctemp}).
So, from (\ref{numero}) it can be seen that
\be
\langle N_{\rm bag} \rangle (\beta, R, -\mu ;-\theta) = - \langle
N_{\rm bag} \rangle (\beta, R, \mu ;\theta) \label{eq-sim2}.\ee
Notice that the total mean fermionic number has the same behavior,
since $N_{Sk} (-\theta) = - N_{Sk} (\theta)$.

From the above relations, one can deduce that
\be
\langle N \rangle (\beta, R, -\mu ;\theta) = 1 - \langle N \rangle
(\beta, R, \mu ;\pi - \theta) \label{eq-sim3} ,\ee allowing to
determine $\langle N \rangle$ for $\mu <0$ from its values for $\mu
> 0$ and $0 \leq \theta \leq \pi$.


\section{Green Function}
\label{sec-green}

In this section we will obtain the Green function in equations
(\ref{eq-eqdiff}),(\ref{eq-bcspatial}),(\ref{eq-bctemp}). In order
to satisfy the ``temporal" boundary conditions, we propose for the
Green function
\be
k(t,\vec{x};t^\prime , \vec{x^\prime}) =R
\sum_{n=-\infty}^{\infty} k_n (\vec{x},\vec{x^\prime})
e^{-i\Omega_n \beta (t-t^\prime)}, \ee
where
\[
\Omega_n = \omega_n - i \mu, \qquad \omega_n = (2n+1)
\frac{\pi}{\beta}.
\]

We adopt the representation for the Euclidean Dirac matrices,
\be
\gamma^0 = i \rho^3 \otimes {\cal I}_2 \qquad \vec{\gamma}=
i\rho^2 \otimes \vec{\sigma} \qquad \gamma^5 = \rho^1 \otimes
{\cal I}_2, \label{gammas} \ee where $\sigma^k$, $\rho^k$ are
Pauli matrices. For each value of $n$, we obtain
\be
\left[ i \omn \rho^3 \otimes {\cal I}_2 \otimes {\cal I}_I -
\rho^2 \otimes \vec{\sigma}\cdot\vec{\nabla}\otimes{\cal I}_I
\right] k_n (\vec{x},\vec{x^\prime}) = \delta^{(3)}
(\vec{x}-\vec{x^\prime}) {\cal I}_2 \otimes {\cal I}_2 \otimes
{\cal I}_I, \ee
with $\omn = \Omega_n R = (2n+1) \pi z - i \mur$,
being $z=RT$ and $\mur=\mu R$ dimensionless variables to be used
along the paper.

We write the Green function as the sum of a particular solution of
the inhomogeneous differential equation plus a solution of the
homogeneous equation,
\be
k_n(\vec{x},\vec{x^\prime}) = k_n^{(0)}(\vec{x},\vec{x^\prime}) +
\tilde{k}_n(\vec{x},\vec{x^\prime}), \ee
with
\be
k_n^{(0)}(\vec{x},\vec{x^\prime}) = \sum_{\ell =0}^{3} \rho^{\ell}
\otimes A^{(\ell,n)} (\vec{x},\vec{x^\prime}), \label{eq-kcero}\ee
\be
\tilde{k}_n(\vec{x},\vec{x^\prime}) = \sum_{\ell =0}^{3}
\rho^{\ell} \otimes a^{(\ell,n)} (\vec{x},\vec{x^\prime}).
\label{eq-kene}\ee

We choose $k_n^{(0)}(\vec{x},\vec{x^\prime})$ so as to give the
contribution of a free gas to the mean fermionic number. The
second term (\ref{eq-kene}) accounts for the correction, due to
the boundary conditions, to the mean fermionic number.

The differential operator (\ref{eq-diffop}) and the boundary
condition operator (\ref{eq-boundop}) are invariant under
transformations in the diagonal subgroup of $SU(2)_{rotations}
\otimes SU(2)_{isospin}$. So, the Green function is block-diagonal
and can be expressed in each invariant subspace in terms of the
eigenstates of $\{ K^2, J^2, L^2,S^2, I^2, K_3\}$, where
$\vec{K}=\vec{L}+\vec{S}+\vec{I}$. For this purpose, we choose the
following basis in the $(k,m)$ subspace:
\[
\begin{array}{ll}
|1\rangle =|k,j=k+\frac12, l=k, m \rangle & |2\rangle
=|k,j=k-\frac12, l=k, m \rangle \\ |3\rangle =|k,j=k+\frac12,
l=k+1, m \rangle & |4\rangle =|k,j=k-\frac12, l=k-1, m \rangle
\end{array}
\]
for $k \neq 0$, and
\[
\begin{array}{l}
|1\rangle_0 =|0,j=\frac12, l=0, 0 \rangle
\\
|3\rangle_0 =|0,j=\frac12, l=1, 0 \rangle
\end{array}
\]
for $k=0$.

The free field contribution to the Green function in the invariant
subspaces labeled by $n,k,m$ is given by
\be
A^{(0)}_{(k,m)} (r,r^\prime) =A^{(1)}_{(k,m)} (r,r^\prime) = 0
\label{eq-a0a1} \ee
\be
A^{(3)}_{(k,m)} (r,r^\prime) = i S_n \omn^2 J_{(k,m)}(r_<)
H_{(k,m)}(r_>) \label{eq-a2}\ee
\be
A^{(2)}_{(k,m)} (r,r^\prime) = -\omn^2 \left\{
\begin{array}{cc}
  J_{(k,m)} (r_<)M_{(k,m)} H_{(k,m)}(r_>) & r^\prime > r \\
  H_{(k,m)} (r_>)M_{(k,m)} J_{(k,m)}(r_<) & r^\prime < r
\end{array}
\right., \label{eq-a3}\ee
where the dependence on $n$ is implicit in the arguments of the spherical
Bessel functions appearing in the expressions for the diagonal matrices
\begin{eqnarray}
J_{(k,m)}(r) & = & {\rm diag} \left( j_k(i S_n \omn r),j_k(i S_n
\omn r),j_{k+1}(i S_n \omn r),j_{k-1}(i S_n \omn r) \right)
\label{eq-jk} \\ J_{(0,0)}(r) & = & {\rm diag} \left( j_0(i S_n
\omn r),j_1(i S_n \omn r)\right) \label{eq-j0}
\\
H_{(k,m)}(r) & = & {\rm diag} \left( h^{(1)}_k(i S_n \omn
r),h^{(1)}_k(i S_n \omn r),h^{(1)}_{k+1}(i S_n \omn
r),h^{(1)}_{k-1}(i S_n \omn r) \right) \label{eq-hk} \\
H_{(0,0)}(r) & = & {\rm diag} \left( h^{(1)}_0(i S_n \omn
r),h^{(1)}_1(i S_n \omn r) \right). \label{eq-h0}
\end{eqnarray}
In previous equations, $S_n$ stands for the sign of $\omega_n$ and
\[
M_{(k,m)}= \left(
\begin{array}{cc}
  0 & {\cal I}_2 \\
  {\cal I}_2 & 0
\end{array}
\right), \qquad M_{(0,0)}= \left(
\begin{array}{cc}
  0 & 1 \\
  1 & 0
\end{array}
\right).
\]

We propose the following Ansatz for the coefficients introduced in
(\ref{eq-kene}) when considering the solution of the homogeneous
differential equation,
\begin{eqnarray}
a^{(0)}_{(k,m)} (r,r^\prime) & = & i S_n \omn^2 J_{(k,m)}(r)
c(0)_{(k,m)} J_{(k,m)} (r^\prime) \nonumber \\ a^{(1)}_{(k,m)}
(r,r^\prime) & = & -i \omn^2 J_{(k,m)}(r)M_{(k,m)} c(0)_{(k,m)}
J_{(k,m)}(r^\prime) \nonumber
\\ a^{(2)}_{(k,m)} (r,r^\prime) & = & - \omn^2 J_{(k,m)}(r)
M_{(k,m)} c(3)_{(k,m)} J_{(k,m)}(r^\prime) \nonumber \\
a^{(3)}_{(k,m)} (r,r^\prime) & = & i S_n \omn^2 J_{(k,m)}(r)
c(3)_{(k,m)} J_{(k,m)}(r^\prime).
\end{eqnarray}

By imposing the boundary conditions, we can determine the unknown
matrices $c(0)_{(k,m)}$ and $c(3)_{(k,m)}$ from the equations
\begin{eqnarray}
A^{(2)}_{(k,m)} (\vec{x},\vec{x^\prime}) & + &a^{(2)}_{(k,m)}
(\vec{x},\vec{x^\prime})+ i \cos\theta \, \left(i \vec{n} \cdot
\vec{\sigma} \otimes {\cal I}_I \right)_{(k,m)} a^{(0)}_{(k,m)}
(\vec{x},\vec{x^\prime}) \nonumber \\ &-& i \sin\theta \, \left( i
\vec{n} \cdot \vec{\sigma} \otimes i \vec{n} \cdot \vec{\tau}
\right)_{(k,m)} a^{(1)}_{(k,m)} (\vec{x},\vec{x^\prime}) = 0,
\end{eqnarray}
\begin{eqnarray}
A^{(3)}_{(k,m)} (\vec{x},\vec{x^\prime}) & + & a^{(3)}_{(k,m)}
(\vec{x},\vec{x^\prime})+ \cos\theta \, \left(i \vec{n} \cdot
\vec{\sigma} \otimes {\cal I}_I \right)_{(k,m)} a^{(1)}_{(k,m)}
(\vec{x},\vec{x^\prime}) \nonumber \\ & - &  \sin\theta \, \left(
i \vec{n} \cdot \vec{\sigma} \otimes i \vec{n} \cdot \vec{\tau}
\right)_{(k,m)} a^{(0)}_{(k,m)} (\vec{x},\vec{x^\prime}) = 0,
\end{eqnarray}
for $|\vec{x}|=1$. The matrices $\left(i \vec{n} \cdot
\vec{\sigma} \otimes {\cal I}_I \right)_{(k,m)}$ and $\left( i
\vec{n} \cdot \vec{\sigma} \otimes i \vec{n} \cdot \vec{\tau}
\right)_{(k,m)}$ are explicitly given in Ref.~\cite{mulders}.

It is straightforward to see that, for $k\neq 0$,
\be
c(0)_{(k,m)} = \left(
\begin{array}{cc}
C^{(k,m)}_0 & 0_{2 \times 2} \\ 0_{2 \times 2} & -C^{(k,m)}_0
\end{array} \right), \qquad
c(3)_{(k,m)} = \left(
\begin{array}{cc}
C^{(k,m)}_{3,+} & 0_{2 \times 2} \\ 0_{2 \times 2} &
C^{(k,m)}_{3,-}
\end{array} \right),
\ee
where
\be
\left[ S_n \cos\theta \, {\cal I}_2 + \frac{\sin\theta}{\nu} E_k +
\frac{\alpha}{2\nu} \sin\theta \, E_k A_k \right] C^{(k,m)}_0 =
D_k, \label{eq-c0k}\ee
\be
C^{(k,m)}_{3,\pm} = - J^{-1}_{\pm} H_{\pm} + \left[ S_n \cos\theta \,
\sigma_3 J^{-1}_\pm J_\mp \pm \frac{\sin\theta}{2 \nu} \sigma_3 +
\frac{\alpha}{2 \nu} \sin\theta \, J^{-1}_{\pm} \sigma_1 J_\pm \right]
C^{(k,m)}_0, \label{eq-c3k}\ee
Here we have introduced $\alpha =
\sqrt{4 \nu^2 -1}$, $\nu=k+1/2$, and
\[
E_k = \left( \begin{array}{cc} e_k & 0 \\ 0 & -e_{k-1} \end{array}
\right), \qquad D_k = \left( \begin{array}{cc} d_k & 0 \\0&
-d_{k-1}
\end{array} \right),\]
\[ A_k = \left( \begin{array}{cc} 0 &
1-\frac{j_{k-1}(i S_n \omn)}{j_{k+1}(i S_n \omn)} \\ - \left(
1-\frac{j_{k+1}(i S_n \omn)}{j_{k-1}(iS_n \omn)} \right) &0
\end{array} \right),
\]
with
\[ e_k = \frac{j_{k+1} (iS_n \omn) j_k (i S_n \omn)}{j_{k+1}^2
(iS_n \omn) - j_k^2 (iS_n \omn)}, \qquad d_k = \frac{j_{k+1} (iS_n
\omn) h_k^{(1)} (i S_n \omn)-j_k (iS_n \omn) h_{k+1}^{(1)} (i S_n
\omn)}{j_{k+1}^2 (iS_n \omn) - j_k^2 (iS_n \omn)}.
\]

Moreover, for $k=0$,
\[
c(0)_{(0,0)} = \left( \begin{array}{cc} C^{(0,0)}_0 & 0 \\ 0 &
-C^{(0,0)}_0 \end{array} \right), \qquad c(3)_{(0,0)} = \left(
\begin{array}{cc} C^{(0,0)}_{3,+} & 0 \\ 0 & C^{(0,0)}_{3,-} \end{array}
\right),
\]
and the coefficients are determined by the equations
\be
\left( S_n \cos\theta \, + 2 \sin\theta e_0 \right) C^{(0,0)}_0 =
d_0, \label{eq-c00}\ee
\be
C^{(0,0)}_{3,+} = \left[ S_n \cos\theta \, (j_0 (i S_n \omn))^{-1}
j_1 (i S_n \omn) + \sin\theta \, \right] C^{(0,0)}_0 - (j_0 (i S_n
\omn))^{-1} h^{(1)}_0 (i S_n \omn) \label{eq-c30p}, \ee
\be
C^{(0,0)}_{3,-} = \left[ S_n \cos\theta \, (j_1 (i S_n \omn))^{-1}
j_0 (i S_n \omn) - \sin\theta \, \right] C^{(0,0)}_0 - (j_1 (i S_n
\omn))^{-1} h^{(1)}_1 (i S_n \omn). \label{eq-ce0m}\ee

The above equations allow for the complete determination of the
Green function. In the next section, we will give the explicit
expressions for those coefficients needed for the evaluation of
the mean fermionic number of the bag.



\section{The mean fermionic number}
\label{sec-constructing}

The mean fermionic number for the bag in the hybrid model can be
written as
\be
\langle N_{\rm bag} \rangle = \langle N_0 \rangle + \langle
\tilde{N}\rangle. \label{numero-splitted} \ee

In equation (\ref{numero-splitted}), $\langle N_0 \rangle$ is the
contribution corresponding to a free fermionic field. It has been
calculated in \cite{gibbons}, to give
\be
\langle N_0 \rangle = N_f \times \frac{4\pi}{9} \left( \mur z^2 +
\frac{\mur^3}{\pi^2} \right), \ee
with the number of flavors $N_f = 2 $ in our case. See Ref.\cite{mu-MIT}
for an evaluation of $\langle N_0 \rangle$ employing techniques similar to
the present paper.

Finally, we will evaluate the second term of
(\ref{numero-splitted}) as
\be
\langle \tilde{N} \rangle = \frac1\beta {\rm Tr} \left[ - i \left(
\gamma^0 \otimes {\cal I}_I \right) \tilde{k} (t,\vec{x};t^\prime,
\vec{x^\prime}) \right]. \ee
It is seen from the Dirac
matrices representation (\ref{gammas}) that only $a^{(3)}
(t,\vec{x};t^\prime,\vec{x^\prime})$ must be considered.

In taking the trace, the integration over angular variables can be
done to obtain
\begin{eqnarray}
\langle \tilde{N} \rangle & = & 2 z \int_0^1 r^2\, dr \,
\int_{\Omega} d\Omega \sum_{n=-\infty}^\infty \left. tr\{ a^{(3,n)}
(\vec{x},\vec{x^\prime})\} \right\rfloor_{\vec{x^\prime}=\vec{x}}
\nonumber \\ & = & 2z \sum_{n=-\infty}^{\infty} i S_n \omn^2
\int_0^1 r^2 \, dr \, \left. \left[  {\rm tr} \left( J_{(0,0)} (r)
c(3)_{(0,0)} J_{(0,0)} (r^\prime) \right)
\phantom{\sum_{m=-k}^{k}}\right. \right. \nonumber \\ & & \left.
\left. + \sum_{k=1}^\infty \sum_{m=-k}^{k} {\rm tr} \left(
J_{(k,m)} (r) c(3)_{(k,m)} J_{(k,m)} (r^\prime) \right) \right]
\right\rfloor_{r^\prime = r} . \label{eq-ntilde}\end{eqnarray}

It is easily seen from Equation (\ref{eq-ntilde}) and from the
definitions (\ref{eq-jk},\ref{eq-j0}) that only the diagonal
elements of $c(3)_{(k,m)}$, $\left[ C_{3,\pm}^{(k,m)}
\right]_{11}$, $\left[ C_{3,\pm}^{(k,m)} \right]_{22}$,
$C_{3,\pm}^{(0,0)}$ are needed. Their explicit evaluation from
equations (\ref{eq-c0k}--\ref{eq-ce0m}) leads to the equalities
\be
\left[ C^{(k,m)}_{3,+} \right]_{11} = \left[ C^{(k,m)}_{3,-}
\right]_{11}, \qquad \left[ C^{(k,m)}_{3,+} \right]_{22} = \left[
C^{(k,m)}_{3,-} \right]_{22}, \qquad C^{(0,0)}_{3,+} =
C^{(k,m)}_{3,-}, \ee allowing to write for $\langle \tilde{N}
\rangle$
\begin{eqnarray}
\langle \tilde{N} \rangle & = & 2z \sum_{n=-\infty}^{\infty} i S_n
\omn^2 \int_0^1 r^2 \, dr \, \left\{   C_{3,+}^{(0,0)} \left[ j_0
(iS_n \omn r) j_0 (iS_n \omn r^\prime) + j_1 (iS_n \omn r) j_1
(iS_n \omn r^\prime) \right] \right.\nonumber \\ & &
 + \sum_{k=1}^\infty \sum_{m=-k}^k \left\{
\left[C_{3,+}^{(k,m)}\right]_{11} \left[ j_k (iS_n \omn r) j_k
(iS_n \omn r^\prime) + j_{k+1} (iS_n \omn r) j_{k+1} (iS_n \omn
r^\prime) \right] \right. \nonumber \\  & & \left. \left. \left. +
\left[C_{3,+}^{(k,m)}\right]_{22} \left[ j_k (iS_n \omn r) j_k
(iS_n \omn r^\prime) + j_{k-1} (iS_n \omn r) j_{k-1} (iS_n \omn
r^\prime) \right] \right\} \right\} \right\rfloor_{r^\prime =r}.
\label{eq-n}
\end{eqnarray}
Notice that we have changed the order in which the series and the integral
are taken.  To properly define the resulting series we will take $r^\prime
= r (1-\epsilon)$, keeping $\epsilon > 0$ up to the end of the
calculation.

In equation (\ref{eq-n}), we can do the $m$-sum, to get a factor
$2 \nu$. We can also rearrange the $k$-sum of the first term to
include the $k=0$ subspace contribution. In this way, we obtain
\begin{eqnarray}
\tilde{N} & = & \lim_{\epsilon \rightarrow 0}
\sum_{n=-\infty}^{\infty} \sum_{k=0}^{\infty} N_{k,m}
(z,\mur,\theta
; \epsilon) \nonumber \\
& = & \lim_{\epsilon \rightarrow 0} \sum_{n=-\infty}^{\infty} \pi
z \omn \sum_{k=0}^{\infty} \left[ 2\nu
\left[C_{3,+}^{(k,m)}\right]_{11} + 2 (\nu+1)
\left[C_{3,+}^{(k+1,m)}\right]_{22} \right] \left( {\rm I}_{\nu} +
{\rm I}_{\nu+1} \right), \label{nro-red}\end{eqnarray}
where we have introduced
\[
{\rm I}_\nu = (1-\epsilon)^{-1/2} \int_0^1 r \, dr \, J_\nu (i S_n
\omn r) J_\nu (iS_n \omn r (1-\epsilon) ),
\]
with $J_\nu (x)$ the cilindrical Bessel functions.

One can see that
\be
{\rm I}_{\nu} + {\rm I}_{\nu +1} =
\frac{(1-\epsilon)^{-3/2}}{\epsilon} \frac{J_\nu (i S_n\omn
(1-\epsilon)) J_\nu (iS_n \omn)}{(i S_n \omn)^2} \left[
d_j^\epsilon - d_j^{\epsilon =0} + \epsilon \left( d_j^{\epsilon
=0} -\nu \right) \right], \label{eq-sum} \ee
where $d_j^\epsilon =
i S_n \omn (1-\epsilon) \left[ \frac{d}{dx} \log J_\nu (x)
\right]_{x=iS_n \omn (1-\epsilon)}$.

The first factor in the $k$-series on the r.h.s. of
(\ref{nro-red}) contains all the dependence of $\langle \tilde{N}
\rangle$ on the chiral angle $\theta(R)$. Using the recurrence
relations for the Bessel functions, we can express this factor in
terms of modified Bessel functions,
\[
2\nu \left[C_{3,+}^{(k,m)}\right]_{11} + 2 (\nu+1)
\left[C_{3,+}^{(k+1,m)}\right]_{22} \]
\be
= \left( \frac{2i}{\pi} \right) \frac{x^2}{(J_\nu (i S_n \omn))^2}
\left[ 2 \nu \frac{M_\nu \bar{N}_\nu + P_\nu
\bar{P}_\nu}{\bar{M}_\nu \bar{N}_\nu + \bar{P}_\nu^2} +
\frac{x^2}{(d_\nu^{\epsilon=0} - \nu)^2} 2 \left(\nu +1 \right)
\frac{\bar{M}_{\nu+1} N_{\nu+1} + P_{\nu+1}
\bar{P}_{\nu+1}}{\bar{M}_{\nu+1} \bar{N}_{\nu+1} +
\bar{P}_{\nu+1}^2} \right], \label{eq-ces}\ee
where
\begin{eqnarray}
M_\nu & = & S_n \cos\theta \left[ u - v + w \right] - \frac{i
\sin\theta}{2 \nu} \frac{\sqrt{1-t^2}}{t} \left[v - 2 t^2 w\right],
\\ N_\nu & = & S_n \cos\theta \left[ -u - v - w
\right] - \frac{i \sin\theta}{2 \nu} \frac{\sqrt{1-t^2}}{t}
\left[v + 2 t^2 w \right], \\ P_\nu & = &- \frac{i
\sin\theta}{2\nu} \frac{\sqrt{1-t^2}}{t} \, v \sqrt{4 \nu^2 -1},
\\ \bar{M}_\nu & = & S_n \cos\theta \left[
\left(d_\nu^{\epsilon=0} - \nu \right)^2 + \nu^2 \left(
\frac{1-t^2}{t^2} \right) \right] - i \sin\theta
\frac{\sqrt{1-t^2}}{t} \left(d_\nu^{\epsilon=0} - \nu \right), \\
\bar{N}_\nu & = &S_n \cos\theta \left[ -\left(d_\nu^{\epsilon=0} +
\nu \right)^2 - \nu^2 \left( \frac{1-t^2}{t^2} \right) \right] - i
\sin\theta \frac{\sqrt{1-t^2}}{t} \left(d_\nu^{\epsilon=0} + \nu
\right),
\\ \bar{P}_\nu & = & - i \sin\theta \frac{\sqrt{1-t^2}}{t} \,
d_\nu^{\epsilon=0} \sqrt{4
\nu^2 -1}.
\end{eqnarray}
Here,
\[
\rho^2 = \nu^2 + x^2, \qquad t = \frac\nu\rho, \qquad x= S_n \omn, \]
\[
u = I_\nu^\prime (S_n \omn) K_\nu^\prime (S_n \omn), \qquad v =
\frac\nu{x} \left( I_\nu (S_n \omn)K_\nu (S_n \omn) \right)^\prime,
\qquad w = \frac{\rho^2}{x^2} I_\nu (S_n \omn)K_\nu (S_n \omn).
\]

\bigskip

The double series (\ref{nro-red}) is not absolutely convergent for
$\epsilon = 0$, so we must keep $\epsilon > 0$ in the general term
up to the end of the calculations. A way to isolate this non
regular behavior consists in the subtraction of the $M$-order
asymptotic (Debye)  expansion \cite{abram} of the general term
$N_{k,m} (z,\mur,\theta ; \epsilon)$, $\Delta_{k,m}^{M}$, as
discussed in Ref.~\cite{mu-MIT}.

So, we can write

\begin{eqnarray}
\langle \tilde{N} \rangle & = & \tilde{N_1} +  \tilde{N_2}
\nonumber \\ &=& \sum_{n=-\infty}^{\infty} \sum_{k=0}^{\infty}
\left[ N_{k,m} (z,\mur,\theta ; \epsilon=0) - \Delta_{k,m}^6
(z,\mur,\theta;\epsilon = 0) \right] \label{eq-debye} \\ && +
\lim_{\epsilon \rightarrow 0} \sum_{n=-\infty}^{\infty}
\sum_{k=0}^{\infty} \Delta_{k,m}^6 (z,\mur,\theta;\epsilon)
\nonumber
\end{eqnarray}

For $M \geq 3$ the $\epsilon \rightarrow 0$ limit can be taken
inside the double sum for $\tilde{N_1}$ . In equation
(\ref{eq-debye}) we have used the expansion up to the order $M=6$
to improve the numerical calculations to be done later. The non
regular behavior of the series for $\langle \tilde{N} \rangle$ is
isolated in the second term $\tilde{N_2}$, which can be studied
analytically.

From equations (\ref{eq-sum},\ref{eq-ces}), by defining
\be
F_1 (z,\mur; \epsilon) = D.D.^{(6)} \left[\frac{I_\nu (S_n \omn
(1-\epsilon)) }{I_\nu (S_n \omn)} \right] \label{eq-efe1}\ee and
\begin{eqnarray}
F_2 (z,\mur,\theta; \epsilon) \nonumber &=&\\ &=& D.D.^{(6)}
\left[ (2 i z) \omn \frac{(1-\epsilon)^{-3/2}}{\epsilon} \left[
d_j^\epsilon - d_j^{\epsilon =0} + \epsilon \left( d_j^{\epsilon
=0} -\nu \right) \right] \right. \nonumber
\\ &  \times & \left. \left[ 2 \nu \frac{M_\nu \bar{N}_\nu + P_\nu
\bar{P}_\nu}{\bar{M}_\nu \bar{N}_\nu + \bar{P}_\nu^2} +
\frac{z^2}{(d_\nu^{\epsilon=0} - \nu)^2} 2 \left(\nu +1 \right)
\frac{\bar{M}_{\nu+1} N_{\nu+1} + P_{\nu+1}
\bar{P}_{\nu+1}}{\bar{M}_{\nu+1} \bar{N}_{\nu+1} +
\bar{P}_{\nu+1}^2} \right] \right] \label{eq-efe2} ,\end{eqnarray}
where $D.D.^{(6)}$ stands for the asymptotic Debye expansion up
to the sixth order, we can write
\be
\Delta_{k,m}^6 (z,\mur,\theta ; \epsilon) = F_1 (z,\mur; \epsilon)
F_2 (z,\mur,\theta; \epsilon) \ee (consistently retaining in this
product terms up to $\rho^{-6}$).

It is easy to see that \cite{mu-MIT}
\be
F_1 (z,\mur; \epsilon) = \exp \left( - \epsilon \frac{\nu}{t}
\right) \left[ 1 + O(\epsilon) \right] \ee

So, we straightforwardly get
\be
\tilde{N_1} = \sum_{n=-\infty}^{\infty} \sum_{k=0}^{\infty} \left[
N_{k,m} (z,\mur,\theta ; \epsilon =0) - F_2
(z,\mur,\theta;\epsilon = 0) \right], \label{eq-ene1}\ee
\be
\tilde{N_2} = \lim_{\epsilon \rightarrow 0}
\sum_{n=_\infty}^{\infty} \sum_{k=0}^{\infty} \exp \left(
-\epsilon \frac\nu{t} \right) F_2 (z,\mur, \theta; \epsilon =0 )
(1 + O(\epsilon)), \label{eq-ene2}\ee
where $F_2 (z,\mur,\theta;
\epsilon =0)$. Employing the asymptotic Debye expansion for Bessel
functions we obtain
\begin{eqnarray}
F_2 (z,\mur, \theta; \epsilon =0 ) & = & (2 i z \omn) \left\{  -
{{{t^5}}\over {{{\nu }^2}}} + {{{t^5}}\over {2\,{{\nu }^3}}}
\left[-3 - t + 5\,{t^2} + 3\,{t^3} - t\,\cos (2\,\theta ) \right]
\begin{array}{c}\\ \\ \end{array} \right. \nonumber \\
& + & {{{t^5}}\over {8\,{{\nu }^4}}} \left[ -6 - 6\,t + 25\,{t^2}
+ 42\,{t^3} + 35\,{t^4} - 48\,{t^5} - 71\,{t^6} \right. \nonumber
\\ & & \left. +
  \left( -6\,t - 14\,{t^2} + 12\,{t^3} + 21\,{t^4} \right) \,\cos (2\,\theta
  ) \right] \nonumber \\
& + & {{{t^5}}\over {32\,{{\nu }^5}}} \left[ -4 - 12\,t +
30\,{t^2} + 212\,{t^3} + 560\,{t^4} - 253\,{t^5} - 2044\,{t^6}
\right. \nonumber \\ & & \left.
-
  690\,{t^7} + 1562\,{t^8} + 852\,{t^9} \right. \nonumber \\ & & \left. +
  \left( -12\,t - 84\,{t^2} - 76\,{t^3} + 406\,{t^4} + 456\,{t^5} - 378\,{t^6} -
     450\,{t^7} \right) \,\cos (2\,\theta ) \right. \nonumber \\ & & \left. +
  \left( 8\,{t^3} - 3\,{t^5} \right) \,\cos (4\,\theta ) \right] \label{eq-desa}\\
& + & {i\over 4}\, {{{t^8}}\over
   {{{\nu }^5}\,
     {{\left( 1 - {t^2} \right) }^{{1\over 2}}}}}
     \cos (\theta )\,{{\sin (\theta )}^3}\, \left[ -8 + 7\,{t^2}\right]\nonumber \\& + &
{{{t^6}}\over {128\,{{\nu }^6}}} \left[ -8 - 20\,t + 504\,{t^2} +
2305\,{t^3} + 414\,{t^4} - 6696\,{t^5} - 17400\,{t^6}
  \right. \nonumber \\ & & \left. - 14025\,{t^7} + 39576\,{t^8} +
50414\,{t^9} - 23856\,{t^{10}} - 32799\,{t^{11}} \right. \nonumber
\\ & & \left.  +
  \left( -8 - 168\,t - 744\,{t^2} + 468\,{t^3} + 7312\,{t^4} + 4842\,{t^5} -
     16700\,{t^6}  \right. \right.\nonumber \\
& & \left. \left. - 15664\,{t^7} + 10800\,{t^8} + 11154\,{t^9}
\right) \,
   \cos (2\,\theta ) \right. \nonumber \\
& & \left. + \left( 48\,{t^2} + 136\,{t^3} - 158\,{t^4} -
270\,{t^5} +
     60\,{t^6} + 99\,{t^7} \right) \,\cos (4\,\theta ) \right] \nonumber \\
& - & \left. {{i}\over 8} \, {{t^8}\over
   {{{\nu }^6}\,{{\left( 1 - {t^2} \right) }^{{1\over 2}}}}} \cos (\theta )\,{\sin (\theta )}^3
\left[ 24 + 68\,t - 91\,{t^2} - 169\,{t^3} + 63\,{t^4} +
100\,{t^5} \right] \right\}\nonumber
\end{eqnarray}

Now, equation (\ref{eq-ene2}) can be expressed in terms of the
double series
\be
s(p,q;\epsilon) = \sum_{k=0}^{\infty} \nu^q {\cal S}_p
(\nu,\epsilon), \qquad t(p,q;\epsilon) = \sum_{k=0}^{\infty} \nu^q
{\cal T}_p (\nu,\epsilon), \label{eq-debye2}\ee
where
\be
{\cal S}_p (\nu,\epsilon) = (-2z) \sum_{n=-\infty}^{\infty}
\frac{i \omn \exp \left( - \epsilon \sqrt{\nu^2 + \omn^2}
\right)}{\left( \nu^2 + \omn^2 \right)^{p/2}}\label{eq-sums}\ee
and
\be
{\cal T}_p (\nu,\epsilon) = (-2z) \sum_{n=-\infty}^{\infty}
\frac{\exp \left( - \epsilon \sqrt{\nu^2 + \omn^2} \right)}{\left(
\nu^2 + \omn^2 \right)^{p/2}}. \label{eq-sumt}\ee

The series ${\cal S}_p$ have been studied in Ref.~\cite{mu-MIT},
and the series ${\cal T}_p$ can be evaluated in a similar way.
Both satisfy the same recursion relations \cite{mu-MIT}. So, they
can be completely determined from the knowledge of ${\cal S}_2
(\nu , \epsilon)$ and ${\cal T}_2 (\nu , \epsilon)$. ${\cal S}_2
(\nu, \epsilon)$ is evaluated in the Appendix~II of Ref.
\cite{mu-MIT}, and ${\cal T}_2 (\nu, \epsilon)$ can be evaluated
following similar steps.

While ${\cal S}_p (\nu, \epsilon)$ is regular at $\epsilon = 0$
for all $p \geq 1$, ${\cal T}_p (\nu, \epsilon)$ is regular at
$\epsilon =0$ only for $p \geq 2$. In both cases they are
exponentially decreasing with $\nu$, making the remaining $k$
series in equations (\ref{eq-debye2}) absolutely convergent, even
for $\epsilon = 0$.

Consequently, it is sufficient to consider the $\epsilon =0$
limit, thus getting the simplified expressions (see
Ref.~\cite{mu-MIT})
\be
{\cal S}_{2 \kappa + 1} (\nu, 0) = \frac{1}{[2 \kappa -1]!!}
\left( - \frac1\nu \frac{\partial}{\partial \nu} \right)^{\kappa }
{\cal S}_{1} (\nu, 0) , \label{s2kp1} \ee
\be
{\cal S}_{2 \kappa} (\nu, 0) = \frac{1}{[2(\kappa -1)]!!} \left( -
\frac1\nu \frac{\partial}{\partial \nu} \right)^{\kappa -1} {\cal
S}_{2} (\nu, 0), \label{s2k} \ee
where
\be
{\cal S}_1 (\nu, 0) = {{2\nu}\over \pi} \int_0^\infty \, du \left[
{1\over{1+e^{-\bar{\mu}/z} e^{{\nu \over z}\sqrt{u^2 +1}}}} -
{1\over{1+e^{+\bar{\mu}/z}e^{{\nu \over z}\sqrt{u^2 +1}}}}\right]
, \label{ese1} \ee
\be
{\cal S}_2 (\nu,0) = {1\over{1+e^{-(\bar{\mu}-\nu)/z}}} -
{1\over{1+e^{(\bar{\mu}+\nu)/z}}} , \label{ese2} \ee and
\be
{\cal T}_{2 \kappa} (\nu, 0) = \frac{1}{[2(\kappa -1)]!!} \left( -
\frac1\nu \frac{\partial}{\partial \nu} \right)^{\kappa -1} {\cal
T}_{2} (\nu, 0), \label{t2k} \ee
\be
{\cal T}_{2 \kappa + 1} (\nu, 0) = \frac{1}{[2 \kappa -1]!!}
\left( - \frac1\nu \frac{\partial}{\partial \nu} \right)^{\kappa
-1} {\cal T}_{3} (\nu, 0) , \label{t2kp1} \ee
with
\be
{\cal T}_2 (\nu,0) = \frac1\nu \left[
{1\over{1+e^{-(\bar{\mu}-\nu)/z}}} +
{1\over{1+e^{(\bar{\mu}+\nu)/z}}} -1 \right], \label{t2} \ee
\be
{\cal T}_3 (\nu, 0) = -{{2}\over \pi} \int_0^\infty \,
\frac{du}{\sqrt{u^2 +1}} \left( \frac1\nu \frac{\partial}{\partial
\nu} \right) \left[ {1\over{1+e^{-\bar{\mu}/z} e^{{\nu \over
z}\sqrt{u^2 +1}}}} + {1\over{1+e^{+\bar{\mu}/z}e^{{\nu \over
z}\sqrt{u^2 +1}}}}\right] - \frac{2}{\pi \nu^2}. \label{t3} \ee

In particular, by inspection of the powers of $\rho$ appearing in
the expression of $F_2 (z,\mur,\theta ;\epsilon=0)$, equation
(\ref{eq-desa}), it can be seen that only $s(p,q,\epsilon =0)$ for
$p\geq5$, and $t(p,q,\epsilon =0)$ for $p\geq8$ are present in
$\tilde{N}_2$.

These results are suitable for performing the numerical
evaluations of $\tilde{N}_2$, which will be described in the next
section.

Numerical evaluations become more difficult when $z$ grows up.
But, in the $z>1$ region an asymptotic approximation valid for
large $z$ applies.

In fact, the series in Eq. (\ref{eq-debye2}) for $p \geq 4$ can be
expressed as
\be
s(p,q; \epsilon = 0) = (-4 z) \frac{1}{p-2}
\frac{\partial}{\partial \mur}\,  \sigma_{p-2,q} \, ,
\label{s-sigma} \ee
\be
t(p,q; \epsilon = 0) = (-4 z)\, \sigma_{p,q} \, , \label{t-sigma}
\ee where
\be
\sigma_{p,q} = \sumk \nu^q \, \sum_{n=0}^{\infty} \rho^{-p}\, .
\ee
As discused in Ref. \cite{mu-MIT}, $\sigma_{p,q}$ can be developed, for
$z>>1$ and $p \geq 2$, as
\[
\sigma_{p,q}  =
\frac12\,  \Re \left[
\left( \frac1{2 \pi z} \right)^{p-q-1}
\frac{\Gamma \left(\frac{p-q-1}{2}\right)
\Gamma \left( \frac{q+1}2 \right)}{\Gamma \left( \frac{p}2 \right)}
\zeta \left( p-q-1 , \frac12 - i \frac{\mur}{2 \pi z} \right) \right. +
\]
\be
\left.
2 \sum_{\ell =0}^{\infty}
\frac{(-1)^\ell}{\ell !} \left(
\frac1{2 \pi z} \right)^{p+2 \ell}
\frac{\Gamma \left( \frac{p + 2\ell}{2}\right)}{\Gamma \left(
\frac{p}2 \right)} \zeta \left( p+2 \ell, \frac12 -i
\frac{\mur}{2 \pi z} \right) \zeta \left( - (q + 2 \ell),
\frac12 \right) \right].
\label{sigma}
\ee

Notice that Eqs. (\ref{sigma},\ref{s-sigma},\ref{t-sigma}) suggest
that $\tilde{N}_1$ is ${\cal O}(z^{-3})$ (since the terms in the
series in Eq. (\ref{eq-ene1}) are ${\cal O}(1/z)^{p-q-4}$ with
$p-q=7$, the next order in the Debye expansion). So, up to this
order, we have
\be
\langle N_{\rm bag} \rangle = {{8\pi}\over{9}} \left( \bar{\mu}
z^2 + \frac{\mur^3}{\pi^2} \right) - {{4\,\mur }\over {3\,\pi }} +
   {{\mur \,
\zeta (3)}\over
     {90\,{{\pi }^3}\,{z^2}}}
\left( 31 - 21\,\cos (2\,\theta ) \right)
+ O(z^{-3})\, .
\ee
The first term in the r.h.s. of this equation is the well known
contribution from the free fermionic field. The remaining terms come from
the large $z$ expansion of $\tilde{N}_2$.
For $\theta=0$ this reproduces the result found in Ref. \cite{mu-MIT} for
the MIT-case.



\section{Numerical evaluations}
\label{sec-results}

In this section we will show our numerical evaluation of $\langle
N \rangle$, for a selected set of values of $z$ and $\theta$, as a
function of $\mur$.

For the double series in equation (\ref{eq-ene1}) we adopt a
cutoff for $n$ and $k$, such that the tail of the series becomes
negligible. Since we have taken the asymptotic expansion up to the
$M=6$ order, the double series are strongly convergent, being the
required maximum values, $n_0$ and $k_0$, small enough for
shortening the numerical calculations.

On the other hand, since we have analytically solved the $n$-sum
in equation (\ref{eq-ene2}), only the remaining $k$-sum in
(\ref{eq-debye2}) must be numerically studied. We impose again a
cutoff to these sums for large values of $k$, so as to also make
negligible the tail of these series.

\begin{figure}
\epsffile{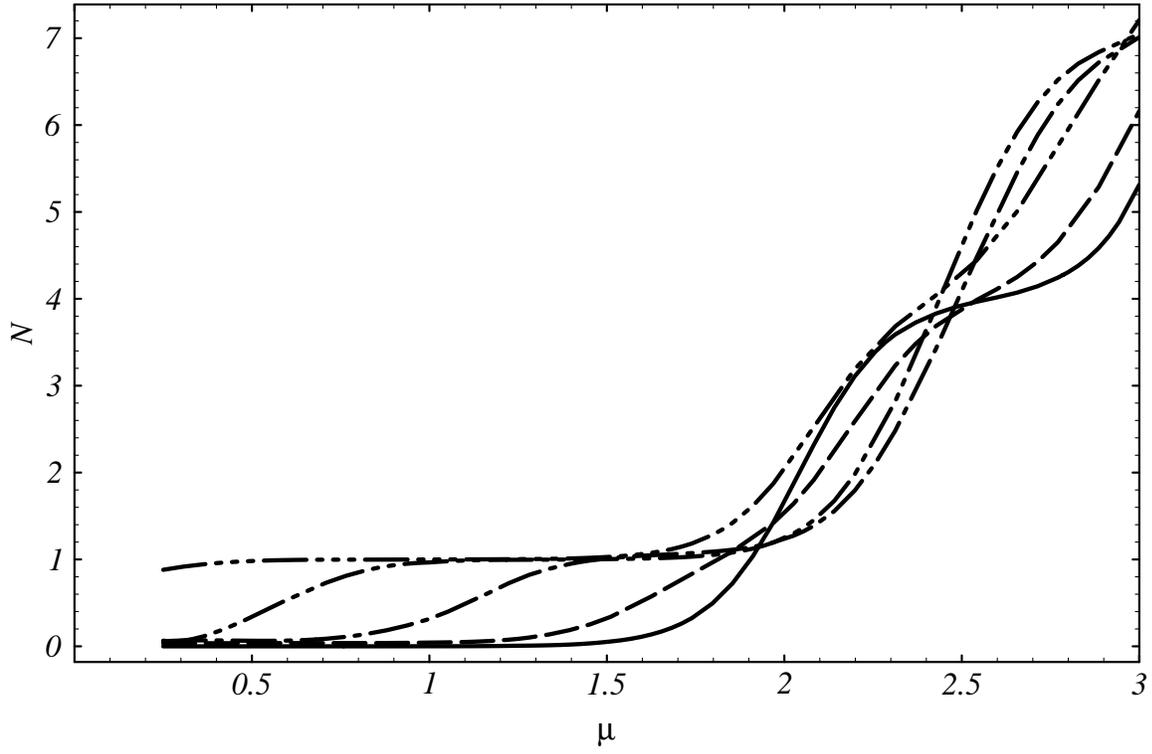} \caption{Fermionic number for
$z=\frac18$,---: $\theta=0$, -- --:
$\theta=\frac{\pi}8$,--$\cdot$--: $\theta=\frac{\pi}{4}$, --$\cdot
\cdot$--: $\theta=\frac{3 \pi}8$,--$\cdot \cdot \cdot$--:
$\theta=\frac{\pi}2$} \label{fig-nume-me}
\end{figure}

\begin{figure}
\epsffile{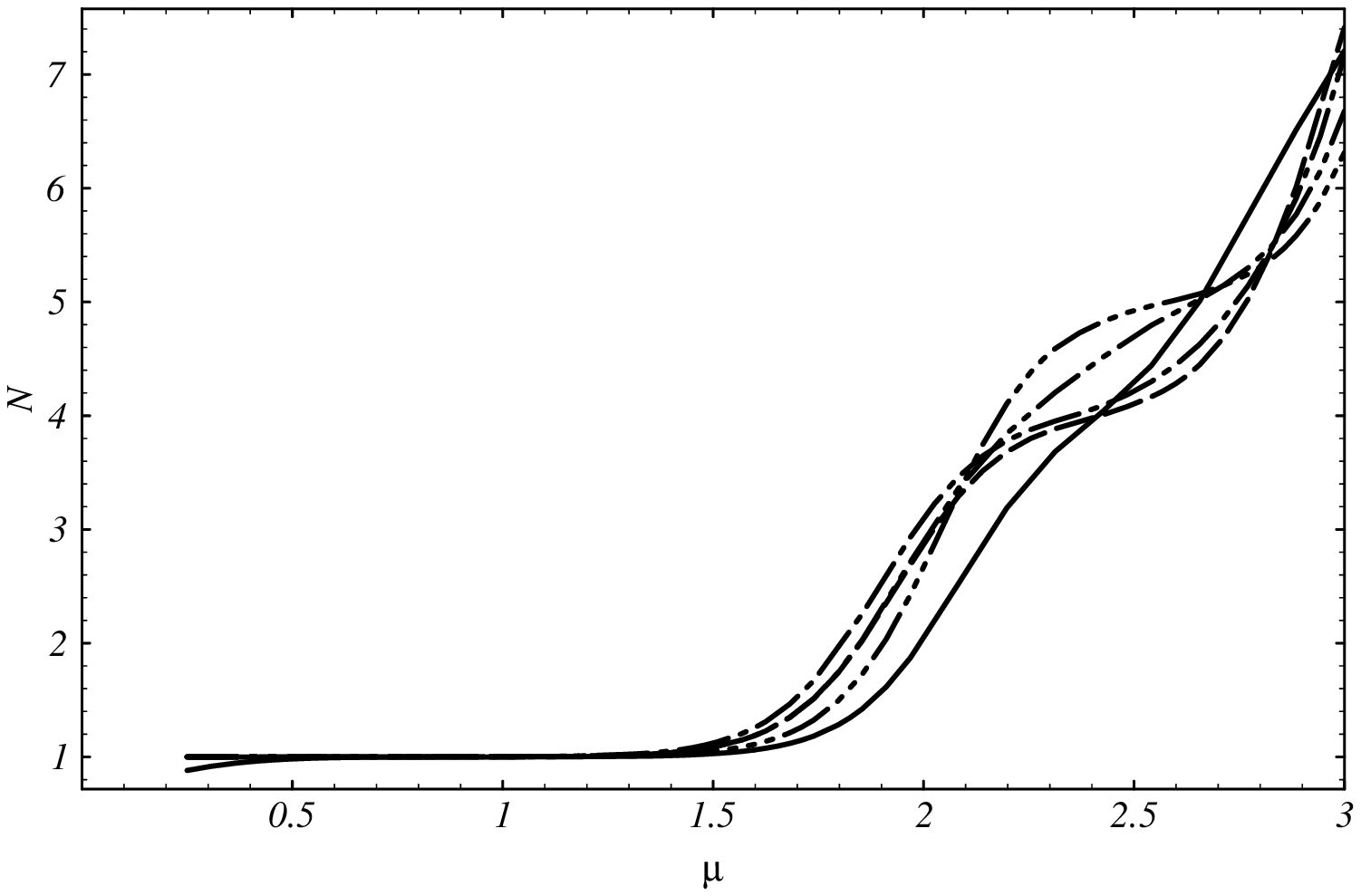} \caption{Fermionic number for
$z=\frac18$,---: $\theta=\frac{\pi}2$, -- --:
$\theta=\frac{5\pi}8$,--$\cdot$--: $\theta=\frac{3\pi}{4}$,
--$\cdot \cdot$--: $\theta=\frac{7 \pi}8$,--$\cdot \cdot \cdot$--:
$\theta=\pi$} \label{fig-nume-ma}
\end{figure}

In Figure~\ref{fig-nume-me} we show the total mean fermionic
number as function of $\mur$, for a fixed value of $z=\frac18$ and
for different values of $\theta \leq \frac{\pi}2$. Similarly,
Figure~\ref{fig-nume-ma} shows the total fermionic number for
$\theta \geq \frac{\pi}{2}$.

The solid line in Fig.~\ref{fig-nume-me} corresponds to the
fermionic number in the MIT bag model ($\theta=0$) \cite{mu-MIT}.
As expected, when $\mur$ approaches the (adimensionalized) energy
of the lowest state of the Dirac Hamiltonian ($RE_0 = 2.04$), the
number jumps up to $\langle N \rangle =4$, which is the
corresponding degeneracy. When $\theta$ grows up, these four
levels split into a singlet and a triplet state \cite{mulders},
the curve having a first step up to $\langle N \rangle = 1$ at a
smaller value of $\mur$. It is worthwhile to remark that $N_{Sk}$
produces a $\theta$-dependent (and $\mur$-independent) shift in
$\langle N \rangle$ (\ref{numero-skyrmion}), which is canceled by
the contribution of the fermionic degrees of freedom, $\langle
N_{\rm bag} \rangle$, giving the expected plateau at $\langle N
\rangle = 1$.

\bigskip

\begin{figure}
\epsffile{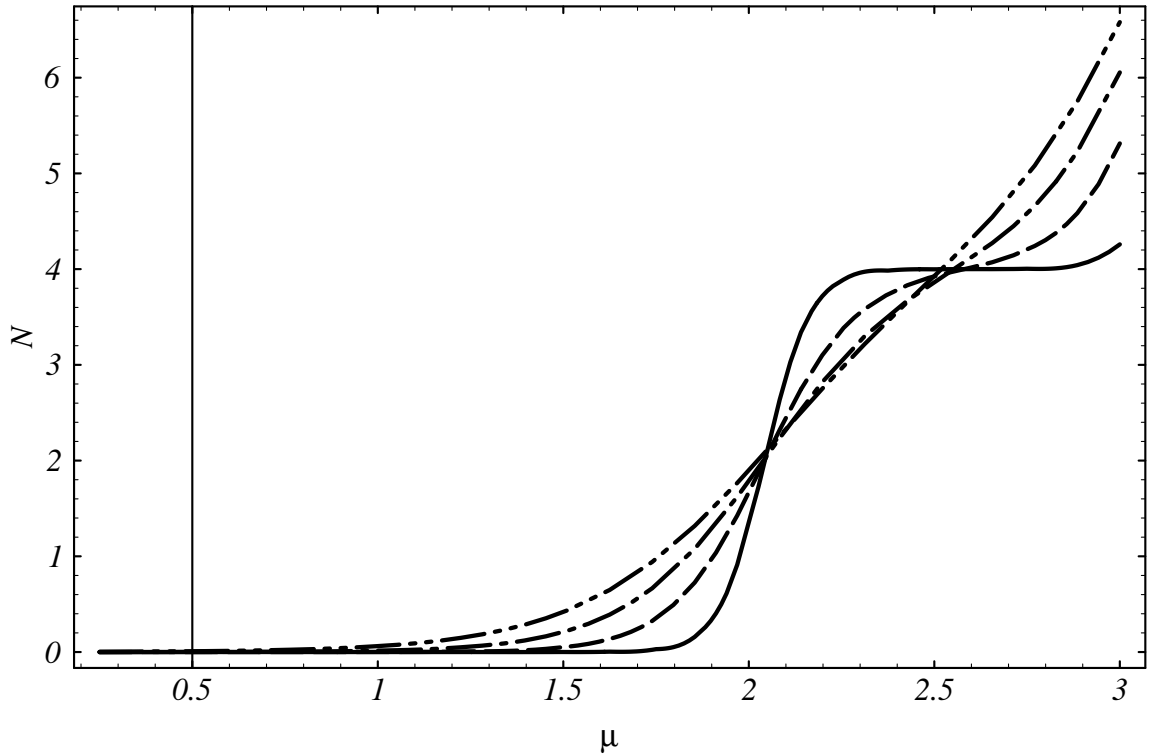} \caption{Fermion number at $\theta=0$;
---:$z=\frac3{50}$, -- --:$z=\frac{1}{8}$,
--$\cdot$--:$z=\frac3{16}$, --$\cdot \cdot$--:$z=\frac1{4}$}
\label{fig-numz_t1}
\end{figure}

\begin{figure}
\epsffile{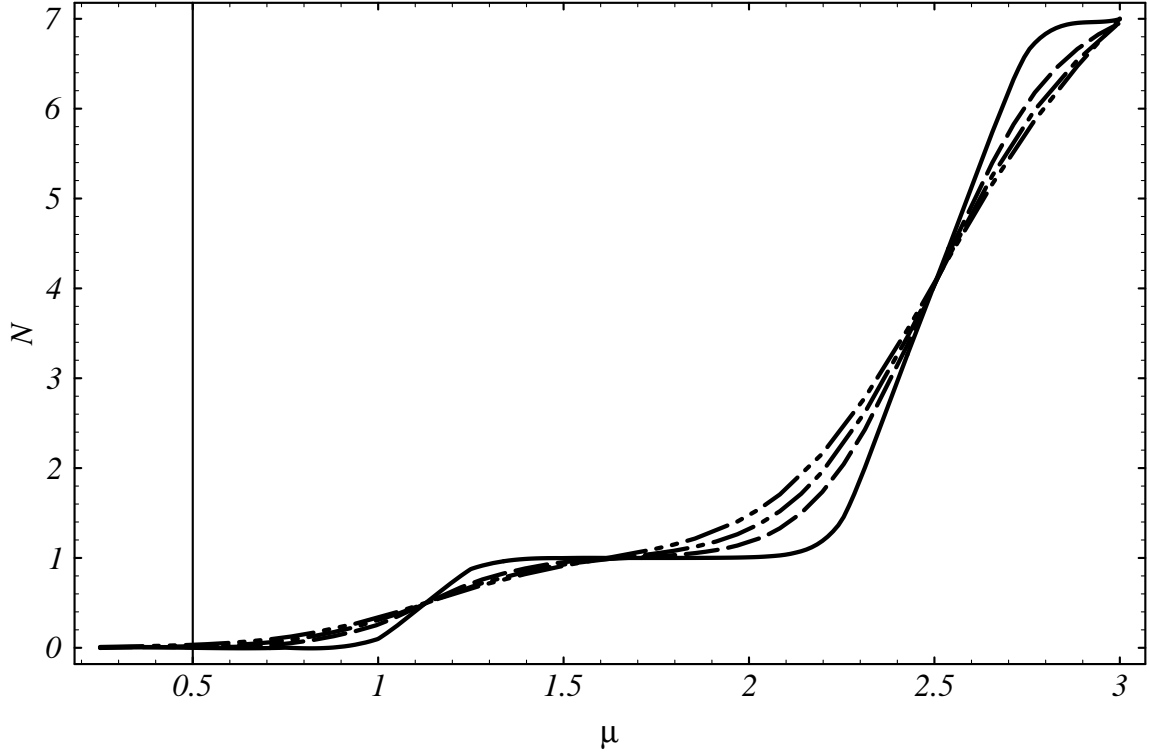} \caption{Fermion number at
$\theta=\frac{\pi}{4}$; ---:$z=\frac3{50}$, -- --:$z=\frac{1}{8}$,
--$\cdot$--:$z=\frac3{16}$, --$\cdot \cdot$--:$z=\frac1{4}$}
\label{fig-numz_t11}
\end{figure}

Finally, we show the total fermionic number for different values
of $z$, for $\theta=0$ in Fig.~\ref{fig-numz_t1} and $\theta=
\pi/4$ in  Fig.~\ref{fig-numz_t11}. These figures show how the way
energy levels are populated smoothes when $z=RT$ grows up .

\section{Conclusions}
\label{sec-conclusion}

In previous sections we evaluated the mean fermionic number for a
hybrid chiral model consisting in a static spherical bag
containing a massless two flavored fermionic field and an exterior
skyrmionic sector. Fermions are constrained to satisfy chiral
boundary conditions depending on the Skyrmion profile $\theta
(R)$ at the the bag surface.

Describing the one-loop Grand partition function through the
functional determinant of the Dirac operator subject to (chiral
bag) spatial and (antiperiodic) temporal boundary conditions, we
were able to express the mean fermionic number in the bag,
$\langle N_{\rm bag} \rangle = {\displaystyle -\frac{\partial
G}{\partial \mu }(\beta ,R,\mu ;\theta)}$, as a trace involving
the Green function of the boundary value problem. This trace has
been expanded as a sum over finite dimensional invariant
subspaces, and evaluated with help of the asymptotic (Debye)
expansion, allowing to show the regularity of the final result.

The CCP requires to complement $\langle N_{\rm bag} \rangle$ with
the contribution from the tail of the Skyrmion, $N_{Sk} (\theta)$.
This introduces a $\theta$-dependent shift in the total mean
fermionic number $\langle N \rangle$, which is exactly compensated
by an opposite one coming from the fermionic field in the bag.
Notice that this behavior of $\langle N_{\rm bag} \rangle$, which
reflects the spectral asymmetry induced by a non vanishing
$\theta$, can not be obtained by considering only a finite number
of eigenvalues of the Dirac operator to approximate the Grand
partition function.

We would like to remark that up to now we have considered the
chiral angle $\theta(R)$ as a free parameter. In fact, the
topological charge contained in the tail of a Skyrmion-like
configuration, Eq.~(\ref{numero-skyrmion}), depends only on
$\theta (R)$ at the surface, and on the boundary condition
$\lim_{r\rightarrow \infty} \theta(r) = 0$
\cite{witten,skyrme,goldstone}.

In the description of a ``baryon" ($\langle N \rangle =1$, since
we are considering only one color) through this two phase hybrid
model, the following picture emerges: for large bag radius
($\theta(R) \rightarrow 0$, i.e. $N_{Sk} \rightarrow 0$), the
chemical potential $\mu$ must be fixed so that $\langle N_{\rm
bag} \rangle = 1$. When $R$ becomes smaller, the chiral angle, and
consequently $N_{Sk}$, rise up. The $\mu$ necessary to get
$\langle N_{\rm bag} \rangle (\beta, R, \mu ; \theta) + N_{Sk}
(\theta)= 1$ diminishes. When the radius is such that $\theta (R)
= \frac{\pi}{2}$ the lowest state of the Dirac Hamiltonian dives
into the Dirac sea. At this point $N_{Sk} = \frac12$, so $\mu$
must be chosen to give $\langle N_{\rm bag} \rangle = \frac12$.
When $R \rightarrow 0$ ($\theta (R) \rightarrow \pi$), the
topological charge reaches one. Then, $\langle N_{\rm bag} \rangle
= 0$, which is compatible with $\mu =0$, as can be seen from
(\ref{eq-sim1},\ref{eq-sim2},\ref{eq-sim3}).

\bigskip

The behavior of hadronic matter at finite temperature and density
has received considerable attention during the last years (See for
example \cite{shuryak}). The main motivation behind this effort is
an attempt to understand properly the generally accepted
possibility of a deconfining phase transition from hadronic matter
to the quark-gluon plasma \cite{smilga}, with applications to
relativistic heavy ion collisions and to the early Universe. In
this framework, the study of the thermodynamical stability of the
hybrid model considered above, through the analysis of the free
energy, may be relevant.

The Gibbs free energy can be obtained by integrating $\langle
N_{\rm bag} \rangle$,
\be
G(\beta, R, \mu; \theta) = F(\beta, R; \theta) - \int_0^{\mu}
\langle N_{\rm bag} \rangle (\beta, R, \mu^\prime; \theta) d
\mu^\prime, \ee
where $F(\beta, R; \theta)$ is the ($\mu=0$) Helmholtz free energy of the
chiral bag, evaluated in \cite{bolsa-quiral}.

When studying the system for a fixed ``baryonic" number the
relation between $\mu$ and $\langle N \rangle$ must be inverted,
which requires more detailed numerical information.

Moreover, the study of the free energy as function of $\beta$ and
$R$ needs the knowledge of the detailed profile $\theta (\beta,
R)$. Even though one can rely on approximated explicit versions as
in Ref.~\cite{atiyah,eskola}, this also would require a greater
numerical computational effort. We will report on that subject
elsewhere.

\acknowledgements

The authors thank E.~M.~Santangelo for many useful discussions and
comments.

This work was supported in part by Fundaci\'{o}n Andes -- Antorchas
under Contract No. C-13398/6, ANPCyT (PICT-0421 and PICT-00039)
and CONICET (Argentina), FONDECyT (Chile) under Grant No. 1980577
and Commission of European Communities under contract
C11*-CT93-0315.


\end{document}